\documentstyle[epsfig]{aipproc}

\begin{document}
\title{The spectral energy distribution of Centaurus A (NGC 5128) \\
       -- A summary of all observations including all CGRO results --}

\author{Helmut Steinle}
\address{Max-Planck-Institut f\"ur extraterrestrische Physik\\
         Postfach 1312, 85748 Garching, Germany}

\maketitle

\begin{abstract}
        Due to its proximity and its importance for the understanding
        of active galaxies and their active nuclei (AGN), Centaurus A has 
        been observed frequently within the last 150 years in all accessible
        wavelength bands. Thus a wealth of data exists which has been 
        compiled into the "NASA Extragalactic Database" (NED). Missing from 
        this compilation are to date almost completely the results of recent
        high energy observations as e.g. obtained by the Compton Gamma Ray 
        Observatory (CGRO). A combination of those recent high energy results
        with all other observations in the NED enables us for the first time to 
        establish the important spectral energy distributions (SEDs) of this 
        closest AGN in different emission states. 
        The combined data have been analyzed to produce SEDs which are from
        contemporaneous data and in addition an attempt was made to derive
        SEDs which are spatially resolved, i.e. SEDs from observations
        which can resolve nucleus and jet in Cen A are treated separately. 
\end{abstract}

\section*{Introduction}

The elliptical galaxy NCG~5128 is the stellar body of the giant
     double radio source Centaurus~A (Cen~A). With a distance of only 3 -- 4 
     Mpc \cite{hui93}, Cen~A is the nearest active galaxy. It contains an
     active nucleus (AGN) and a jet with a large inclination ($\sim 70^{\circ}$)
     to the line-of-sight which is detected in all wavelength bands where the 
     spatial resolution is sufficient. Cen~A belongs to the Fanaroff-Riley
     type I galaxies and is often also classified as a Seyfert 2.\\
     Its proximity makes Cen~A uniquely observable among such objects and it
     is a very well studied and frequently observed galaxy in all 
     wavelength bands. Its emission is detected from radio to high-energy 
     gamma-rays \cite{johnson97,israel98,clay94} making it 
     the only radio galaxy detected in MeV gamma-rays. All other AGN 
     detected in MeV gamma-rays (and identified) are blazars \cite{collmar99} where the jet is aligned almost parallel to our line-of-sight.    
     Because Cen~A is seen under a much larger angle, it may be a 
     representataive of the many other "normal" active 
     galaxies which are just too far away to be detected with present day
     instruments sensitive in gamma rays.\\
     To study the global spectral energy distribution (SED) of Cen~A over
     all available frequencies (energies) gives insight into the 
     emission processes in AGN and may even provide hints to the source of the
     cosmic diffuse background at gamma-ray energies.   

\section*{Data}

     All available data have been combined into Fig. 1. About 40 \% of the
     data (122 data points as of March 2001) are from the NASA Extragalactic 
     Database (NED) \cite{ned}. This data base is rather complete up to about 
     $10^{18}$ Hz (4 keV (EINSTEIN data)), but lacks all high energy
     observations. Thus all available data from the Compton Gamma-Ray
     Observatory (CGRO) taken during its more than 9 years of operation and 
     very-high-energy (VHE) observations summarized by \cite{clay94} have 
     been added to the data set so that almost 300 data points are now 
     available. 

\section*{Spatial Resolution of the Observations}

     Because Cen~A is so close, the galaxy can be resolved into the nucleus and
     the outer regions, including the jet, with many of the instruments used. 
     However, especially the instruments observing in the gamma-ray regime lack 
     this resolution (OSSE several degrees, COMPTEL few degrees, EGRET half 
     degree; all on board CGRO).
     Many authors, however, assume that the high energy 
     emission observed can only originate in the nucleus and that emission from 
     the jet is not visible if the object is viewed far from the jet axis (as 
     is the case in Cen~A with a viewing angle of $\sim 70^{\circ}$). 
     Therefore, other than in cases where the spatial resolution of the 
     observations was unknown, the CGRO data are included in the plots of the
     nuclear data, but they are marked differenly. 

\section*{Temporal Resolution of the Observations}

     Centaurus A is known to be a highly variable object in all wavelength
     bands and to show distict emission states 
     \cite{bond96,baity81,turner97,steinle98} 
     Therefore it is very 
     important to measure complete SEDs simultaneous at a given time and
     in the different emission states. This is mandatory to avoid confusion in 
     the interpretation of the data and difficulties when models are fitted to 
     the data. However, simultaneous multiwavelength observations covering a 
     large interval in frequencies have so far only been organized once in 1995
     \cite{steinle99} when Cen A was observed in a low emission state. 
     All other data have been taken at random times.\\
     A problem related to the variability is the fact, that especially 
     observations with low sensitivity instruments require very often long 
     integration times, which are much longer than the typical time scales for 
     the Cen~A variability. Gamma-ray measurements by the instruments on board 
     CGRO lasted typically several weeks, whereas Cen~A is known to be variable 
     in the adjacent hard X-ray band on time scales of less than a day.  

\section*{Classification of Data}

\begin{figure}[b!] 
\centerline{\epsfig{file=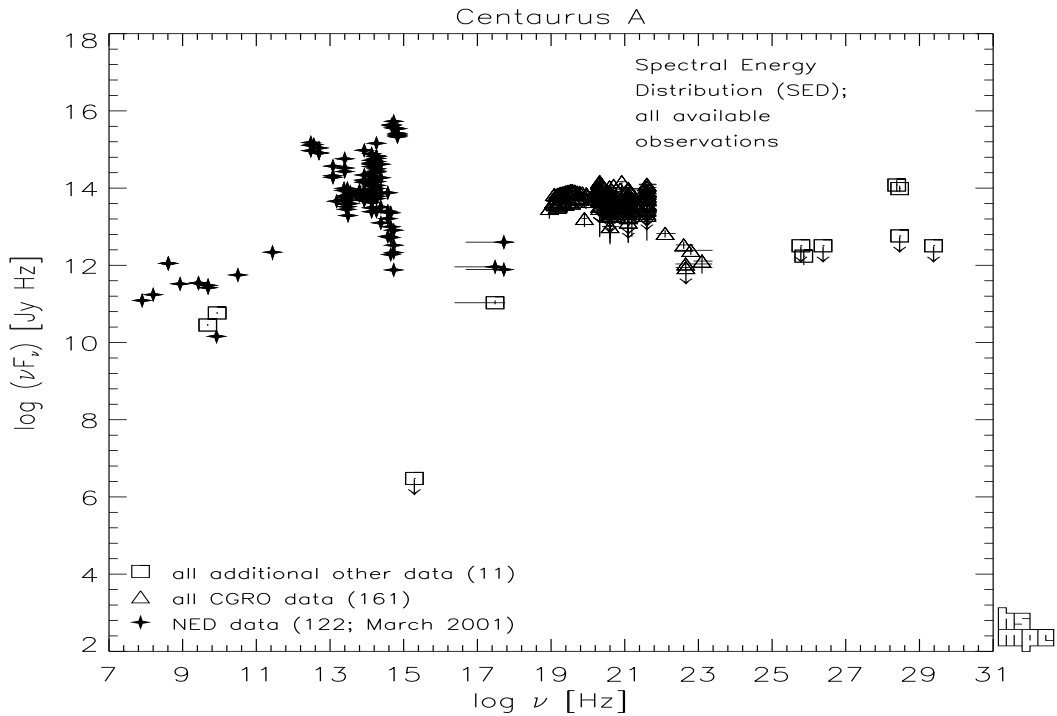,width=5.5in,height=3.8in,angle=0}}
\vspace{10pt}
\caption{All available data from NED, CGRO and other observations}
\label{fig1}
\end{figure}
     To help to draw conclusions from this large collection of data (Fig. 1) 
     in a reasonable manner, the data have been separated into the following 
     groups:

     \begin{itemize}
      \item according to spatial resolution:
      \begin{itemize}
       \item spatial resolution unknown 
       \item spatial resolution not sufficient to resolve Cen A 
       \item spatial resolution sufficient to observe the nucleus alone or if
             it is reasonably assumed that the emission is from the nucleus 
             only (Fig. 2)
      \end{itemize}
      \item simultaneous observations: 
      \begin{itemize}
       \item observations without exact date or averages of many observations 
       \item simultaneous observations (including "long" observations of low
             sensitivity instruments as e.g. the gamma-ray instruments on
             board CGRO) (Fig. 2)
      \end{itemize}
     \end{itemize}

     Due to the limited information available on some of the data, an analysis 
     of the original papers reporting the observations is necessary and will be
     conducted in the future.

\section*{Results}

\begin{figure} 
\centerline{\epsfig{file=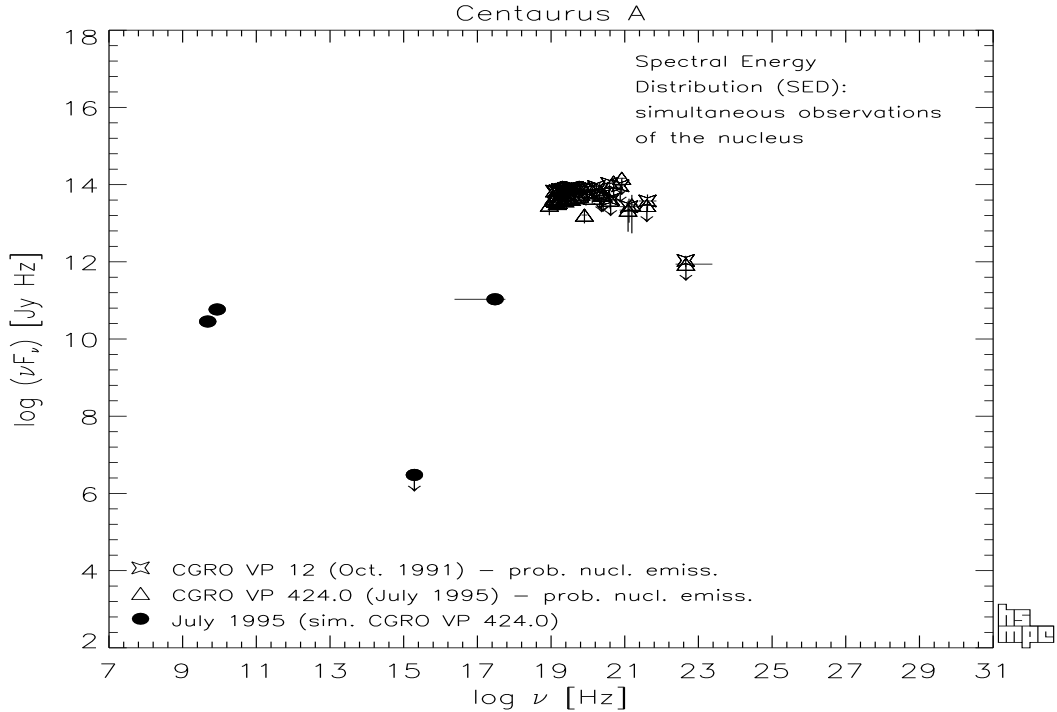,width=5.5in,height=3.8in,angle=0}}
\vspace{10pt}
\caption{All simultaneous observation of the nucleus (see text).}
\label{fig2}
\end{figure}

      In Fig. 1, the global structure of the spectral energy distribution
      of Cen~A shows the typical two "bumps" which are usually (for Blazars
      see e.g. \cite{urry98}) attributed to synchrotron emission and 
      Compton-scattering for the low frequency (here: $\sim 10^{14}$ Hz) and 
      high frequency peak (here: $\sim 10^{20}$ Hz) respectively. Possibly 
      there are two more "bumps" at very low and very high frequencies. However
      at low frequencies, this impression may be caused by the scatter of
      the data points. The simultaneous observations in 1995 (see Fig. 2)
      do not support the presence of a "bump" at lower frequencies. On the
      high frequency side of the SED, the two detections of Cen~A at 
      $\sim 10^{28.5}$ Hz have been questioned and await confirmation from
      instruments available soon with much more sensitivity.

      As one can see from the figures, despite the huge amount of
      data, only few data sets end up in the most interesting Figure 2.
      This shows dramatically the lack of coordinated simultaneus observations,
      an omission which hardly can be corrected, as the Compton Gamma-Ray
      Observatory, which covered a very important spectral region with its
      four instruments and which contributed many important measurements, was
      eliminated before further scheduled coordinated observations had
      taken place. No near-term future gamma-ray instrument will be able to
      close this gap in energy and in observational data.

\section*{Acknowledgements}

I thank R.C. Hartman and F.W. Stecker for pointing out an error which I made in
the plots of the EGRET data in the first version of the paper. This research 
has made use of the NASA/IPAC Extragalactic Database (NED) which is operated by 
the Jet Propulsion Laboratory, California Institute of Technology, under 
contract with the National Aeronautics and Space Administration.

\end{document}